\documentclass[
 reprint,
 amsmath,amssymb,
 aps,
]{revtex4-2}
\usepackage{graphicx}% Include figure files
\usepackage{dcolumn}% Align table columns on decimal point
\usepackage{bm}% bold math
\newcommand{\eg}{\textit{e}.\textit{g}.~}
\newcommand{\program}[1]{\textsc{#1}} % typesetting names of programs

\begin{document}

\title{Constraint for a light charged Higgs boson and its neutral partners\\
from top quark pairs at the LHC}

\author{ChunHao Fu}
\author{Jun Gao}
\affiliation{
    INPAC, Shanghai Key Laboratory for Particle Physics and
    Cosmology \& School of Physics and Astronomy,
    Shanghai Jiao Tong University, Shanghai 200240, China
}
\affiliation{
    Key Laboratory for Particle Astrophysics and Cosmology (MOE),
    Shanghai 200240, China
}

%\date{\today}

\begin{abstract}

The charged Higgs boson plays an essential role in distinguishing between a wide variety of standard model extensions with multiple Higgs doublets, and has been searched for in various collider experiments. This paper expands our previous work to a broader Higgs mass space with discussions on subsequent issues. We study the prospect of a light charged Higgs boson, produced by top quark pairs at the Large Hadron Collider (LHC), and decaying into a $W$ boson (can be off shell) and a pair of bottom quarks via on-shell production of an intermediate neutral Higgs boson. We reinterpret the cross sections of $WWbb\bar{b}\bar{b}$ final states measured by the ATLAS collaboration at LHC 13\,TeV in the presence of the decay chain: $t \rightarrow H^+ b, H^+ \rightarrow W^+ H_i, H_i \rightarrow b \bar{b}$, and H.c., where $H_i$ is a neutral Higgs boson variably lighter than the charged Higgs boson. We find improved agreements with the data and obtain limits on the total branching ratio of the aforementioned decay chain. The limits impose the strongest constraints on the parameter space of type-I two-Higgs-doublet model for most Higgs masses sampled when $H_i$ is the $CP$-odd Higgs boson $A$. We also calculate potential constraints with pseudodata in high-luminosity runs of the LHC.

\end{abstract}

\maketitle

\section{Introduction}

While the Higgs boson, the unique fundamental scalar particle in the standard model (SM) of particle physics, has been hunted down in the ATLAS and CMS experiments at the Large Hadron Collider (LHC) at CERN \cite{ATLAS2012, CMS2012}, the detail of the scalar sector is not yet fully revealed to us eleven years after, and the possibility of a larger and more complex scalar sector is still appealing for many reasons, \eg the supersymmetry \cite{Haber1985}. The simplest but well-motivated extension of the SM scalar sector is the two-Higgs-doublet model (2HDM) \cite{Branco2012, Lee1973}, which features a pair of charged Higgs bosons and three neutral Higgs bosons. The charged Higgs boson is important for identifying various SM extensions from the nature; direct searches on it have been carried out at LEP \cite{LEP2013}, Tevatron \cite{CDF2011}, and now LHC \cite{CMS2020b, ATLAS2021}, while the signals can be affected by possible undiscovered neutral Higgs bosons. Hopefully, the improved sensitivity at the LHC Run 3 will take us further in discovering or denying the charged Higgs boson and its possible non-SM companions.

A light charged Higgs boson has a mass smaller than the mass difference between a top quark and a bottom quark, and can therefore be generated from the decay $t \rightarrow H^+ b$ (and H.c.), which benefits from the large production cross sections of the top quark pair at the LHC. Signals of the light charged Higgs boson can be examined in various final states \cite{Branco2012, Bahl2021, Li2022}. In our previous work \cite{prev_work}, we have studied the $W^\pm b \bar{b}$ final states, which can usually be generated following two patterns: (a) $H^+ \rightarrow t^{(\ast)} \bar{b} \rightarrow W^+ b \bar{b}$, and (b) $H^+ \rightarrow W^+ H_i \rightarrow W^+ b \bar{b}$, where $H_i$ can be any neutral Higgs boson in the model. These patterns are exactly included in 2HDM \cite{Branco2012}, where $H_i$ can be the non-SM scalar $H_\text{non-SM}$ which we refer to as $H$ afterwards, and the pseudoscalar $A$; the SM-like scalar $H_\text{SM}$ is unable to participate in such decay at the alignment limit \cite{Bahl2021}. A few simplifications were made for pattern (b) in 2HDM in our previous work \cite{prev_work}, most significantly being $m_{H^\pm} - m_A = 85\,\text{GeV}$ and $m_H > m_{H^\pm}$. These simplifications are reverted or discussed in further detail in this work. The $W^+ b \bar{b}$ channel has shown its power in exploring the parameter space of 2HDM in many theoretical studies \cite{Akeroyd1999, Arhrib2017, Sanyal2019, Bahl2021, Arhrib2021, Cheung2022, Slabospitskii2022}.

In this paper, we utilize a measurement on inclusive and differential fiducial cross sections of final states composed of two $W$ bosons and four bottom quarks, performed by the ATLAS collaboration at LHC 13\,TeV with an integrated luminosity of 36.1\,fb$^{-1}$ \cite{ATLAS2019}. We reinterpret the data in a two-dimensional Higgs mass space featuring the charged Higgs boson and a neutral Higgs boson $H_i$ with a mass from the threshold of bottom quark pair production to the vicinity of the SM Higgs mass. In this context, after the production of a top quark pair, one of the top quarks may decay as $t \rightarrow H^+ b$ which is followed by $H^+ \rightarrow W^+ H_i \rightarrow W^+ b \bar{b}$, while the other one follows the SM decay $\bar{t} \rightarrow W^- \bar{b}$.

In Section~\ref{sec:generic}: We introduce our data selection and the methodology of our calculation, and then we perform signal-only likelihood tests for the signal strength of the charged-Higgs physics, $B(t \rightarrow H^+ b, H^+ \rightarrow W^+ H_i, H_i \rightarrow b \bar{b})$. We find improved agreements with the LHC data, and set upper limits on the signal strength. Making the neutral Higgs mass a variable shows us a much more complete aspect of the limit contours in the Higgs mass space. We also make explicit comparison between the sensitivity of different data sets to the signal strength.

In Section~\ref{sec:2HDM-I}: We show the general constraints on the possible Higgs mass hierarchies in type-I 2HDM for reference to the Higgs mass space we study. We translate our signal constraints into strong constraints on the parameter $\tan\beta$ of type-I 2HDM with the mass hierarchy $m_H > m_{H^\pm} > m_A$, while paying attention to alternative decay channels contributing to the same final states. The relevant differences between the pseudoscalar $A$ and the scalar $H$ in type-I 2HDM are discussed, and we find scenarios where $m_H < m_{H^\pm}$ can share the same constraint result under certain conditions. We also discuss the potential of future high-luminosity data in both Section \ref{sec:generic} and \ref{sec:2HDM-I}.

\section{\label{sec:generic}Theory and Signal Constraints}

The ATLAS collaboration measured the $t\bar{t}$ production in association with additional $b$-jets, in final states including two $W$ bosons and four bottom quarks. The fiducial cross section measurements were performed in a di-lepton channel where one of the $W$ bosons decays into an electron while the other into a muon, and a lepton-plus-jets channel where one of the $W$ bosons decays into an electron or muon while the other into jets \cite{ATLAS2019}. The $W$ bosons can decay into electrons and muons either in a direct manner or via intermediate tauons. The results have been unfolded to particle level, identifying final states with at least four $b$-jets or at least three $b$-jets (since some $b$ quarks can be out of experimental acceptance). Detailed definitions of the fiducial region can be found in Ref.~\cite{ATLAS2019} and are also implemented in the public \program{Rivet} \cite{Rivet3} analysis routine. We note that there is another measurement on similar final states performed by the CMS collaboration \cite{CMS2020}; however, it cannot be used in this work since it requires reconstructions of the top quarks following the SM decay mode.

Theoretical predictions on the binned cross sections in the presence of a light charged Higgs boson and an even lighter neutral Higgs boson can be calculated as
\begin{eqnarray}
    \sigma^\text{bin}_\text{pre} &=&
    \sigma^\text{bin}_\text{SM} + \sigma^\text{bin}_{H^+}
    \nonumber\\ &:=&
    \sigma_\text{SM}(t\bar{t}b\bar{b}) \epsilon^\text{bin}_\text{SM} + 2B^\text{sig}_{H^+} \sigma_\text{SM}(t\bar{t}) \epsilon^\text{bin}_{H^+},
    \label{eqn:x_sec}
\end{eqnarray}
when the branching ratio of the non-SM decay mode $t \rightarrow H^+ b$ is small. $\sigma_\text{SM}(X)$ denotes the SM cross section of the QCD production of $X$; $\epsilon^\text{bin}_{\text{SM(}H^+\text{)}}$ is the particle-level experimental efficiency for the prescribed kinematic bin of the SM(charged-Higgs) process, with SM branching ratios of the $W$ boson decay. Other SM processes contributing to the same final states are already subtracted from the experimental data. $B^\text{sig}_{H^+}$, representing the signal strength of the charged-Higgs physics, is defined as the branching ratio
\begin{eqnarray}
    B^\text{sig}_{H^+} &:=&
    B(t \rightarrow H^+ b, H^+ \rightarrow W^+ H_i, H_i \rightarrow b \bar{b})
    \nonumber\\ &=&
    B(t \rightarrow H^+ b) \times B(H^+ \rightarrow W^{+(\ast)} H_i)
    \nonumber\\ &&
    \times B(H_i \rightarrow b \bar{b}),
    \label{eqn:B_sig}
\end{eqnarray}
where the first line specifies a process with its $s$-channels (leading to a symmetry factor of 2 in Eq.~(\ref{eqn:x_sec})); validity of the narrow-width approximation followed and on-shellness of the neutral Higgs boson $H_i$ are accepted within the mass space we consider. In the following discussions and calculations, we take the pseudoscalar $A$ in 2HDM as a main example of $H_i$. For type-I 2HDM at or close to the alignment limit and
\begin{eqnarray}
    &
    \tan\beta > 1,
    \quad
    m_{H^\pm} = 100 \text{--} 160 \,\text{GeV},
    \nonumber\\&
    m_A \leqslant 110 \,\text{GeV},
    \quad
    m_H > m_{H^\pm},
    \label{eqn:param_space}
\end{eqnarray}
we confirm that the decay width of $H^+$ is at most $\sim 10^{-1}$\,GeV, and the decay width of $A$ (at most $\sim 10^{-3}$\,GeV) is much smaller than that of $W^+$, using the \program{ScannerS-2} program \cite{ScannerS-2,ScannerS-1}. For the scalar $H$ and \mbox{type-II,X,Y} 2HDMs, there are calculations indicating similar conclusions \cite{Aoki2009}. Apart from the process included by $B^\text{sig}_{H^+}$, (a) the non-resonant production of $H^+$ and (b) an alternative decay mode $H^+ \rightarrow t^{(\ast)} \bar{b}$ can also contribute to the same final states, and we do not include them for simplicity. Contribution (a), for reference, is generally around 10$\%$ of the contribution from the resonant decay of a top quark pair at $m_{H^\pm} \approx 160$\,GeV and even less at smaller $m_{H^\pm}$, in type-I,II 2HDMs \cite{Degrande2017}. Contribution (b) is assumed to be insignificant, and we will discuss the validity of this assumption for type-I 2HDM in Section~\ref{sec:2HDM-I}. We treat $B^\text{sig}_{H^+}$ as an input variable to Eq.~(\ref{eqn:x_sec}), and derive the efficiency $\epsilon^\text{bin}_{H^+}$ from Monte Carlo (MC) simulations of generic 2HDM. The efficiency at particle level represents the size of the detected part, in a specific fiducial channel, of the overall normalized phase space distribution, and can therefore be calculated as
\begin{equation}
    \epsilon^\text{bin} = \sigma^\text{bin}_\text{fid}\,/\,\sigma_\text{MC},
    \label{eqn:efficiency_formula}
\end{equation}
where $\sigma^\text{bin}_\text{fid}$ is the binned cross sections that fall into the fiducial region in the simulated process, and $\sigma_\text{MC}$ is the actual inclusive cross section of the simulated process. We set the decay widths of $H^+$ and $A$ in MC simulations as values small enough to guarantee the on-shellness of the Higgs bosons, and therefore $\epsilon^\text{bin}_{H^+}$ depends only on the masses of the Higgs bosons at leading order.

We perform a survey on the inclusive cross sections and various differential fiducial cross sections measured by ATLAS and select three data sets. The first is the inclusive fiducial cross sections in the di-lepton channel and the lepton-plus-jets channel with at least three or four $b$-jets (totalling four bins). The other two are the normalized distributions of the invariant mass of the two closest $b$-jets in angular distance, $m_{bb}^{\Delta \text{min}}$, with at least three $b$-jets in the di-lepton channel and at least four $b$-jets in the lepton-plus-jets channel respectively. Each normalized distribution was divided into five bins, and we drop the last bin for selecting independent bins. The light neutral Higgs boson produces a pair of bottom quarks with a small invariant mass, which makes these two bottom quarks tend to have a small angular distance. Therefore, the charged-Higgs physics generally enhances differential cross sections at small $m_{bb}^{\Delta \text{min}}$, and distributions of $m_{bb}^{\Delta \text{min}}$ can be quite sensitive to this change. This intuitive conclusion could be violated if the neutral Higgs boson is not very light (rather above 70\,GeV).

We generate event samples with MC simulations in \program{MadGraph5\_aMC@NLO 3.4.2} \cite{MadGraph_ref} followed by parton showering (PS) and hadronizations with \program{PYTHIA 8.306} \cite{PYTHIA8.3} in the four-flavor number scheme (4FS), and analyze the events with the public routine of the ATLAS analysis in \program{Rivet} \cite{Rivet3}. We use CT18 parton distribution functions (PDF) \cite{Hou2021} and a top(bottom) quark pole mass of 172.5(4.7)\,GeV in simulations, and set the default renormalization and factorization scales to the sum of transverse energy of all final states divided by two. For MC simulations of the charged-Higgs process in generic 2HDM, we use the \program{2HDM\_NLO} model from \program{FeynRules} \cite{FeynRules_ref}. The efficiency $\epsilon^\text{bin}_{H^+}$ is calculated using event samples generated at leading order in QCD matched with PS. We set the total cross section of SM $t\bar{t}$ production to 838.5\,pb at LHC 13\,TeV, calculated with \program{Top++ 2.0} \cite{Toppp2_ref1, Toppp2_ref2} at next-to-next-to-leading order (NNLO) and next-to-next-to-leading logarithmic accuracy in QCD.

We have used the general procedure to calculate SM predictions at next-to-leading order (NLO) in QCD and found the result agrees well with the theoretical predictions in the ATLAS analysis \cite{ATLAS2019}. In the remaining part of our study, we instead use the SM predictions in the ATLAS analysis as described below for its comprehensive estimation of uncertainties. For the inclusive fiducial cross sections, there are four predictions generally agreeing with each other, and we use the theoretical prediction from \program{SHERPA 2.2} \cite{SHERPA_ref} at NLO+PS in 4FS, with uncertainties obtained by varying the renormalization and factorization scales by factors of 0.5 and 2.0 and including PDF uncertainties from NNPDF3.0 NNLO PDFs \cite{NNPDF2015}. For the normalized distributions of $m_{bb}^{\Delta \text{min}}$, we take four different predictions from \program{POWHEG} \cite{POWHEG_ref1, POWHEG_ref2, POWHEG_ref3} +\program{PYTHIA8}: a prediction in 4FS for $t\bar{t}b\bar{b}$ production \cite{Jezo2018}, and three predictions with different tunes of the programs \cite{ATLAS2016} in the five-flavor number scheme for $t\bar{t}$ production \cite{Frixione2007} (additional $b$ quarks are generated from PS). We take the mean of the four predictions as the prediction used, thus keeping the normalization, and the standard deviation of the four predictions as the uncertainty. The PDF uncertainties are negligible for normalized distributions and are thus not included. We have checked that evaluating the prediction and uncertainty in a different manner or with different MC results can hardly impact on our final results.

\begin{figure}
    \includegraphics{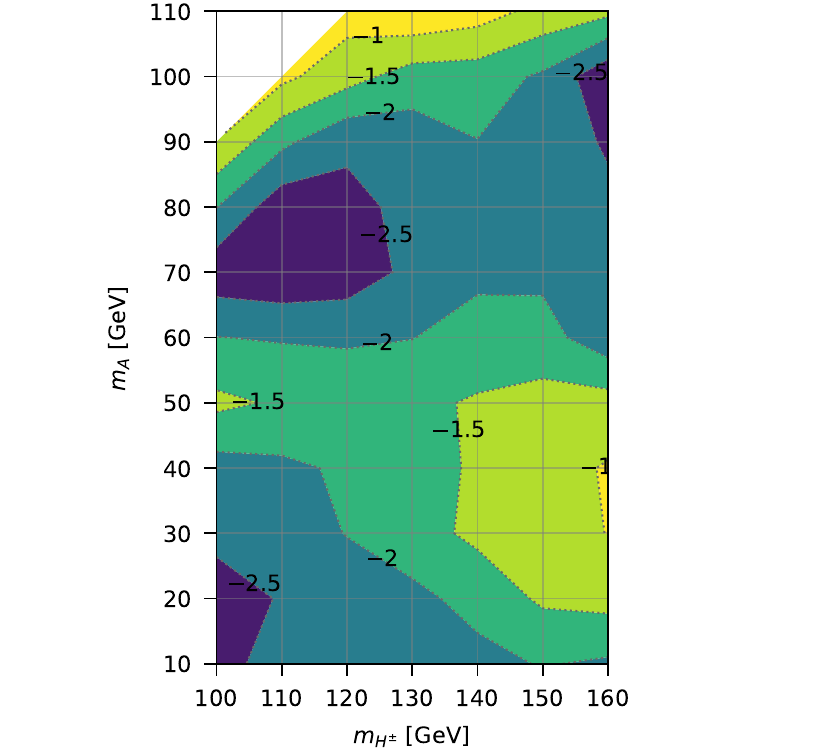}
    \caption{Intuitive contours\footnote{Contours in this work are all plotted using \program{Matplotlib} \cite{Matplotlib}, which implements a marching squares algorithm to compute contour locations based on the sample points.} of $\chi^2_\text{best}$ on the $(m_{H^\pm}, m_A)$ plane, minus the $\chi^2$ of the pure-SM prediction. Note that the best-fitted $B^\text{sig}_{H^+}$ at each point varies with the point.}
    \label{fig:chi2_contours}
\end{figure}
We use an interval of 10\,GeV to sample the Higgs mass space specified in Eq.~(\ref{eqn:param_space}). For each Higgs mass point, the log-likelihood combining all selected data sets together is calculated as
\begin{eqnarray}
    \chi^2(B^\text{sig}_{H^+}, m_{H^\pm}, m_A)
    = \sum_{i = 1}^{N_\text{bin}}
        \frac{(\sigma^{\,i}_\text{pre} - \sigma^{\,i}_\text{exp})^2}
        {\delta^{\,i}_\text{pre}{^2} + \delta^{\,i}_\text{exp}{^2}},
    \label{eqn:chi_square}
\end{eqnarray}
where $\sigma^{\,i}_\text{pre}$ is the theoretical prediction for the $i$-th bin calculated as in Eq.~(\ref{eqn:x_sec}), with an error of $\delta^{\,i}_\text{pre}$; $\sigma^{\,i}_\text{exp}$ is the central value of the measurement in the $i$-th bin, with its statistical error and systematic error combined to be of $\delta^{\,i}_\text{exp}$. The data sets we select include $N_\text{bin} = 12$ uncorrelated kinematic bins. Let $\chi^2_\text{best}$ be the smallest $\chi^2$ possible when varying $B^\text{sig}_{H^+}$ at each Higgs mass point, we then plot $\chi^2_\text{best}$ contours on the $(m_{H^\pm}, m_A)$ plane in Fig.~\ref{fig:chi2_contours}. The pure-SM value $\chi^2(B^\text{sig}_{H^+} = 0) = 7.0$ is subtracted from the contours. The overall best fit among the sample points is found at $(m_{H^\pm} = 100\,\text{GeV}, m_A = 20\,\text{GeV})$ with $B^\text{sig}_{H^+} = 0.54\%$, where $\chi^2$ is lowered by 3.0 units compared to the SM case. We find generally moderate improvements on description of the ATLAS data; the improvements are especially attributed to enhancements to the inclusive fiducial cross sections compared to SM, as is visualized in our previous work \cite{prev_work}.

\begin{figure}
    \includegraphics{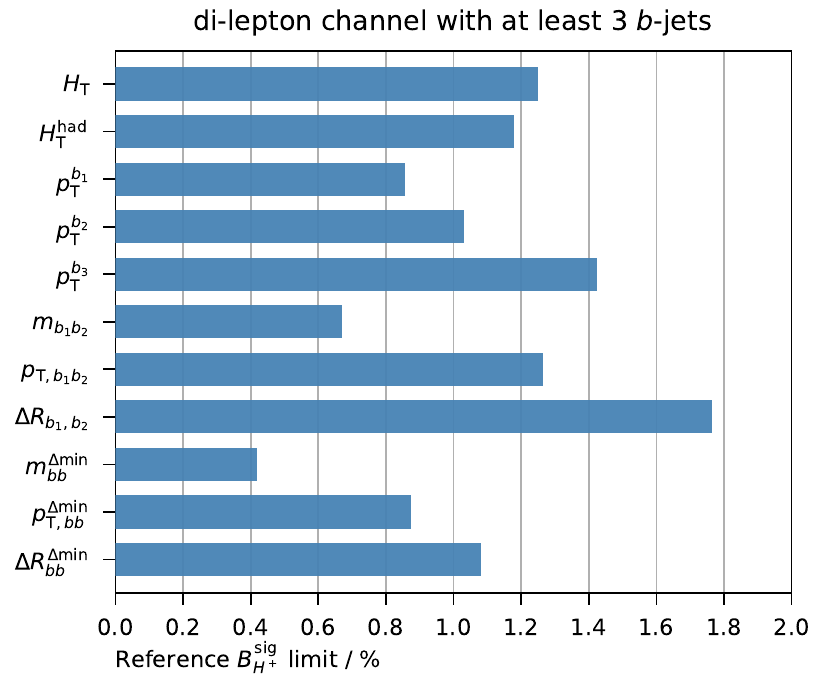}
    \includegraphics{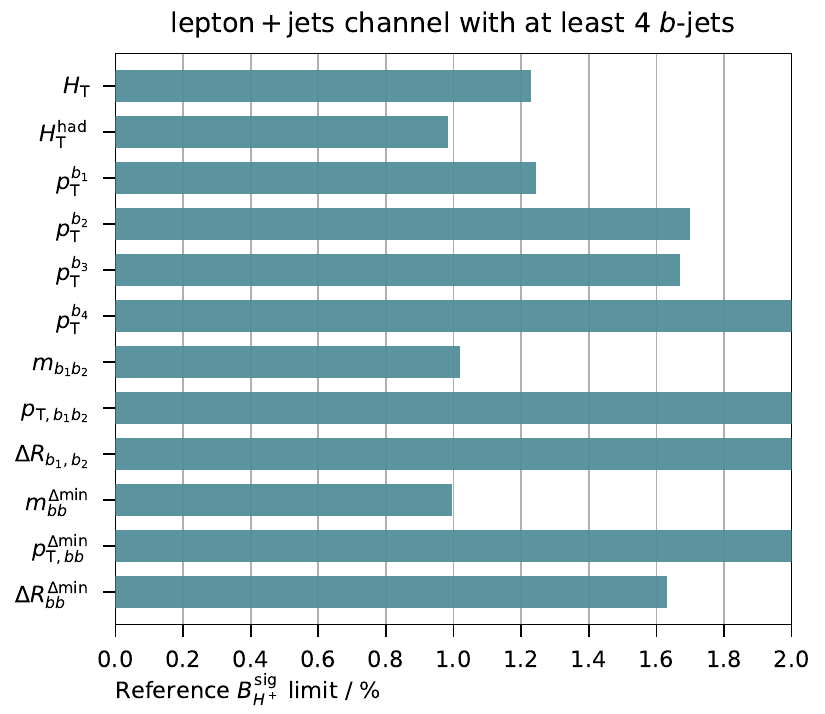}
    \caption{Reference limits on $B^\text{sig}_{H^+}$ for each data set of normalized distribution in the di-lepton channel with at least three $b$-jets and the lepton-plus-jets channel with at least four $b$-jets respectively. The kinematic variables include: (1) the scalar sum of the transverse momenta $p_\text{T}$ of the lepton(s) and jets in the events ($H_\text{T}$) and that of only jets in the events ($H^\text{had}_\text{T}$), (2) $p_\text{T}$ of the $i$-th highest-$p_\text{T}$ $b$-jet ($p_\text{T}^{b_i}$), (3) the invariant mass, $p_\text{T}$, and angular distance of the first and second highest-$p_\text{T}$ $b$-jets ($m_{b_1 b_2}$, $p_{\text{T,}b_1 b_2}$, and $\Delta R_{b_1 b_2}$), and those of the two closest $b$-jets in angular distance ($m_{bb}^{\Delta \text{min}}$, $p_{\text{T,}bb}^{\Delta \text{min}}$, and $\Delta R_{bb}^{\Delta \text{min}}$). A smaller limit value corresponds to a greater average sensitivity of a data set to the decay channel studied in this work.}
    \label{fig:data_set_survey}
\end{figure}
\begin{figure*}
    \begin{minipage}[t]{0.49\linewidth}
        \includegraphics{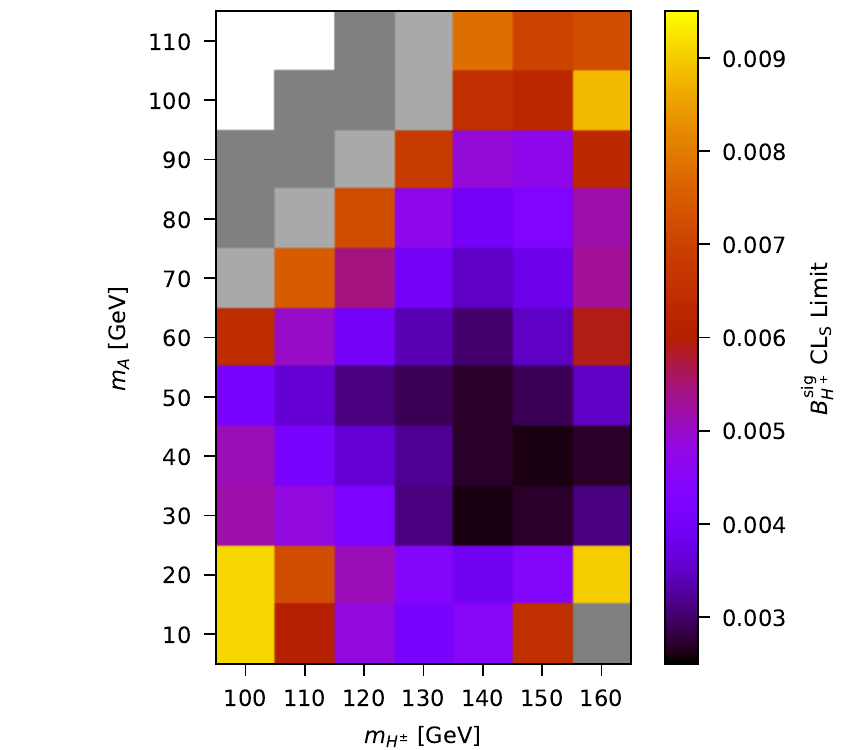}
        \caption{95\% $\text{CL}_\text{s}$ upper limits on $B^\text{sig}_{H^+}$ on the $(m_{H^\pm}, m_A)$ plane. Each tile corresponds to a sample point which is located at the tile's center. The lighter grey tiles represent limits between 1\% and 1.5\%, and the darker grey tiles represent limits over 1.5\%.}
        \label{fig:BR_dist}
    \end{minipage}
    \hfill
    \begin{minipage}[t]{0.49\linewidth}
        \includegraphics{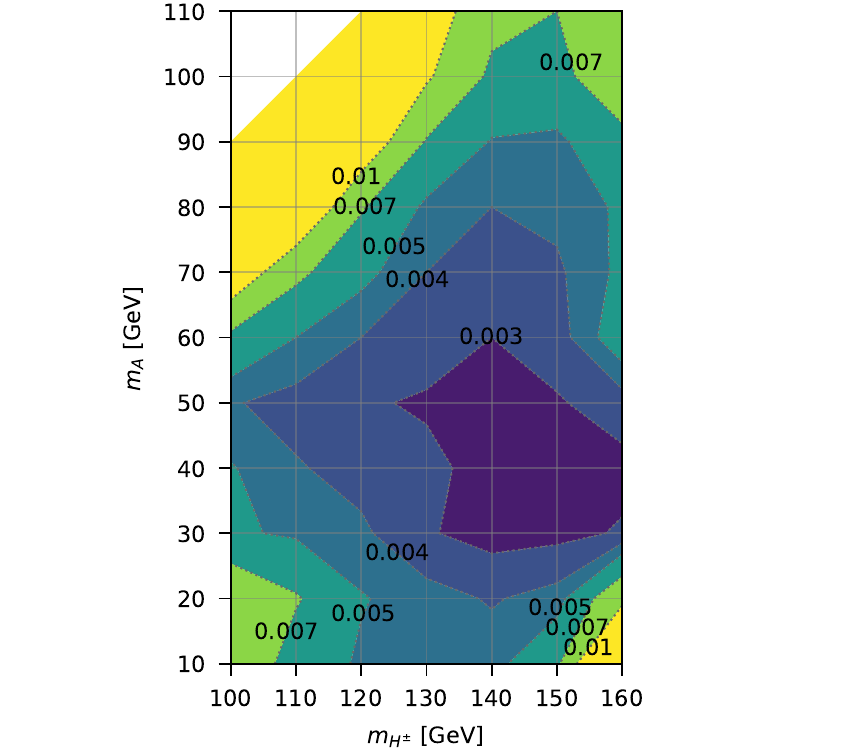}
        \caption{Intuitive contours of the 95\% $\text{CL}_\text{s}$ upper limits on $B^\text{sig}_{H^+}$ on the $(m_{H^\pm}, m_A)$ plane.}
        \label{fig:BR_contours}
    \end{minipage}
\end{figure*}
\begin{figure}
    \includegraphics{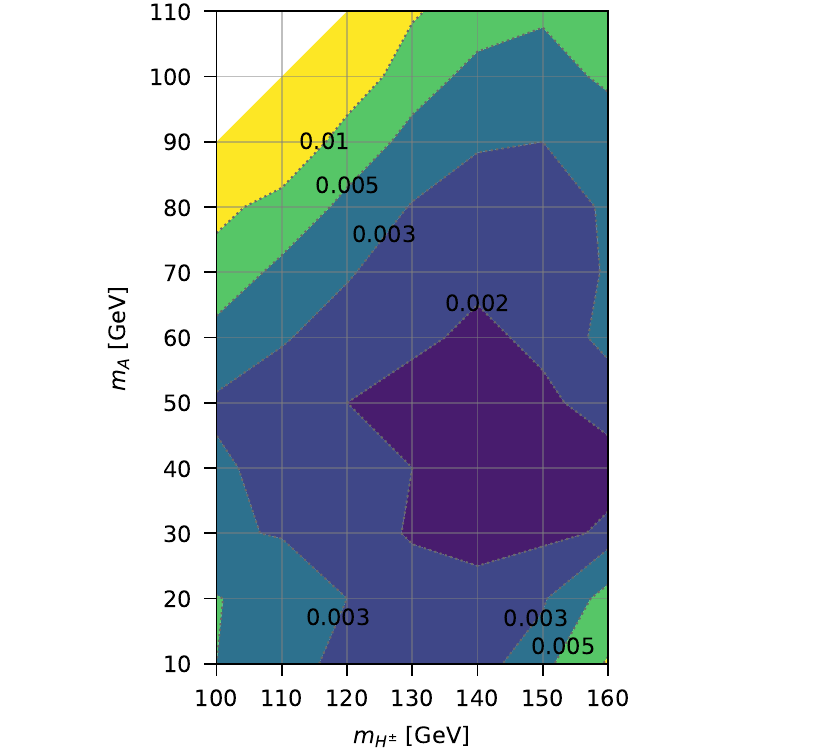}
    \caption{Intuitive contours of the 95\% $\text{CL}_\text{s}$ upper limits on $B^\text{sig}_{H^+}$ that can be potentially reached with SM-like high-luminosity data in the future, on the $(m_{H^\pm}, m_A)$ plane. The statistical uncertainties are reduced to complete insignificance at around 900\,fb$^{-1}$, where the systematic uncertainties prevent further lowering of the signal limits.}
    \label{fig:Projection_BR_contours}
\end{figure}
Based on the values of the $\chi^2$ function, we can use the $\text{CL}_\text{s}$ method \cite{Read2002} to deduce upper limits on $B^\text{sig}_{H^+}$ for fixed $(m_{H^\pm}, m_A)$. The upper limits at a signal-only confidence level of $(1 - \alpha')$ are calculated as
\begin{equation}
    \hat{B} + \delta_B \,\Phi^{-1}[1 - \alpha' \,\Phi(\hat{B} / \delta_B)],
    \label{eqn:CLs_formula}
\end{equation}
where $\hat{B}$ is the best-fitted $B^\text{sig}_{H^+}$ at $(m_{H^\pm}, m_A)$ (corresponding to the central values of the observation), $\delta_B$ is an uncertainty estimated by increasing $B^\text{sig}_{H^+}$ from $\hat{B}$ until $\Delta\chi^2 = 1$ (corresponding to the combined uncertainties of the observation and the prediction), and $\Phi$ is the cumulative distribution function of the standard normal distribution. We note that only the two sets of normalized distributions of $m_{bb}^{\Delta \text{min}}$ are directly included in deducing our final limits when calculating $\chi^2$ (thus $N_\text{bin} = 8$), for the selected distributions have generally smaller relative uncertainties compared with the inclusive cross sections.

Before introducing our final result of the $\text{CL}_\text{s}$ limits on $B^\text{sig}_{H^+}$, we present a result of using $\text{CL}_\text{s}$ limits to quantify the sensitivity of different data sets to the charged-Higgs signal. The ATLAS analysis has presented 25 sets of relative differential cross section data \cite{ATLAS2019}, and we calculate 95\% $\text{CL}_\text{s}$ upper limits on $B^\text{sig}_{H^+}$ at all sample Higgs mass points using 23 of them one by one. The two distributions of the number of $b$-jets are not included as they depend on details of the model due to their $\sigma(t \bar{t})$ normalization. We note the ATLAS analysis has shown that various SM MC predictions in general describe the differential cross section data well within the experimental uncertainties \cite{ATLAS2019}. To simplify the calculation, we set the central values and uncertainties of the SM-predicted distributions involved as corresponding observed central values and zeros respectively. We take the median of the $B^\text{sig}_{H^+}$ limits calculated at all sample points as a reference value which conversely represents a data set's sensitivity to $B^\text{sig}_{H^+}$, and plot these reference limits in Fig.~\ref{fig:data_set_survey}. The normalized distribution of $m_{bb}^{\Delta \text{min}}$ gives clearly the best limit in the di-lepton channel and almost the best limit in the lepton-plus-jets channel, consolidating our previously mentioned conclusion that the normalized distributions of $m_{bb}^{\Delta \text{min}}$ are most sensitive to the charged-Higgs signal. We must note, however, the normalized distributions of $m_{bb}^{\Delta \text{min}}$ are most sensitive in an average sense, and in the lepton-plus-jets channel with at least four $b$-jets, the normalized distribution of $m_{b_1 b_2}$ shows a much greater sensitivity for a relatively heavy light neutral Higgs boson, specifically for sample points with $(m_{H^\pm} \leqslant 130\,\text{GeV}, m_A \geqslant 70\,\text{GeV})$. We expect to see stronger limits on $B^\text{sig}_{H^+}$ especially in this region of Higgs masses in future multivariate analyses.

We now plot the true 95\% $\text{CL}_\text{s}$ upper limits on $B^\text{sig}_{H^+}$ at the sample Higgs mass points in Fig.~\ref{fig:BR_dist}, and the corresponding contours in Fig.~\ref{fig:BR_contours}. The most strict limit found among the sample points is 0.26\%, at two diagonally adjacent points $(m_{H^\pm} = 140\,\text{GeV}, m_A = 30\,\text{GeV})$ and $(m_{H^\pm} = 150\,\text{GeV}, m_A = 40\,\text{GeV})$. The result expands our previous work \cite{prev_work} to the $(m_{H^\pm}, m_A)$ plane, showing the background of the previous $m_{H^\pm} - m_A = 85\,\text{GeV}$ result. We note again that the specific neutral Higgs boson $A$ is replaceable, and the result should also apply to scenarios where $A$ is replaced with any neutral Higgs boson that couples to fermions in a Yukawa way, \eg the non-SM $CP$-even Higgs $H$ in 2HDM, since the efficiency $\epsilon^\text{bin}_{H^+}$ in Eq.~(\ref{eqn:x_sec}) should hardly change when varying parameters other than Higgs masses at leading order.

We can take the view that in future high-luminosity data, the central values may coincide with the pure-SM prediction. Then Eq.~(\ref{eqn:CLs_formula}), whose first term becomes zero and second term receives most contributions from kinematic bins most sensitive to $B^\text{sig}_{H^+}$, approximates the potential upper limits on $B^\text{sig}_{H^+}$ for high-luminosity data within the limits of current systematic uncertainties. To simulate a high luminosity, we lower the statistical uncertainties and the SM theoretical uncertainties of the normalized $m_{bb}^{\Delta \text{min}}$ distributions by 80\% and 50\% respectively, making the combined non-systematic uncertainties be generally around 10\% of the corresponding systematic uncertainties for $B^\text{sig}_{H^+}$ values covered by the calculation. We plot the contours of the resulting 95\% $\text{CL}_\text{s}$ upper limits on $B^\text{sig}_{H^+}$ in Fig.~\ref{fig:Projection_BR_contours}. In this high-luminosity scenario, the observed limits shown in Fig.~\ref{fig:BR_contours} can be lowered by 32--45\% at most sample Higgs mass points (less for $m_A = 110\,\text{GeV}$). Lowering the statistical uncertainties by 80\% intuitively requires an integrated luminosity of around 900\,fb$^{-1}$, while the high-luminosity LHC program is expected to reach 3000\,fb$^{-1}$. It is worth expecting further increased precision especially in the $B^\text{sig}_{H^+}$-sensitive distributions of $m_{bb}^{\Delta \text{min}}$ in future experimental data.

\section{\label{sec:2HDM-I}Constraints on type-I 2HDM}

Constraints on the signal strength $B^\text{sig}_{H^+}$ can be translated into constraints on the parameter space of various models at fixed Higgs masses. Here we discuss constraints on the type-I 2HDM, which is less constrained by direct searches at the LHC among the common 2HDMs \cite{Sanyal2019}.

\begin{figure}
    \includegraphics{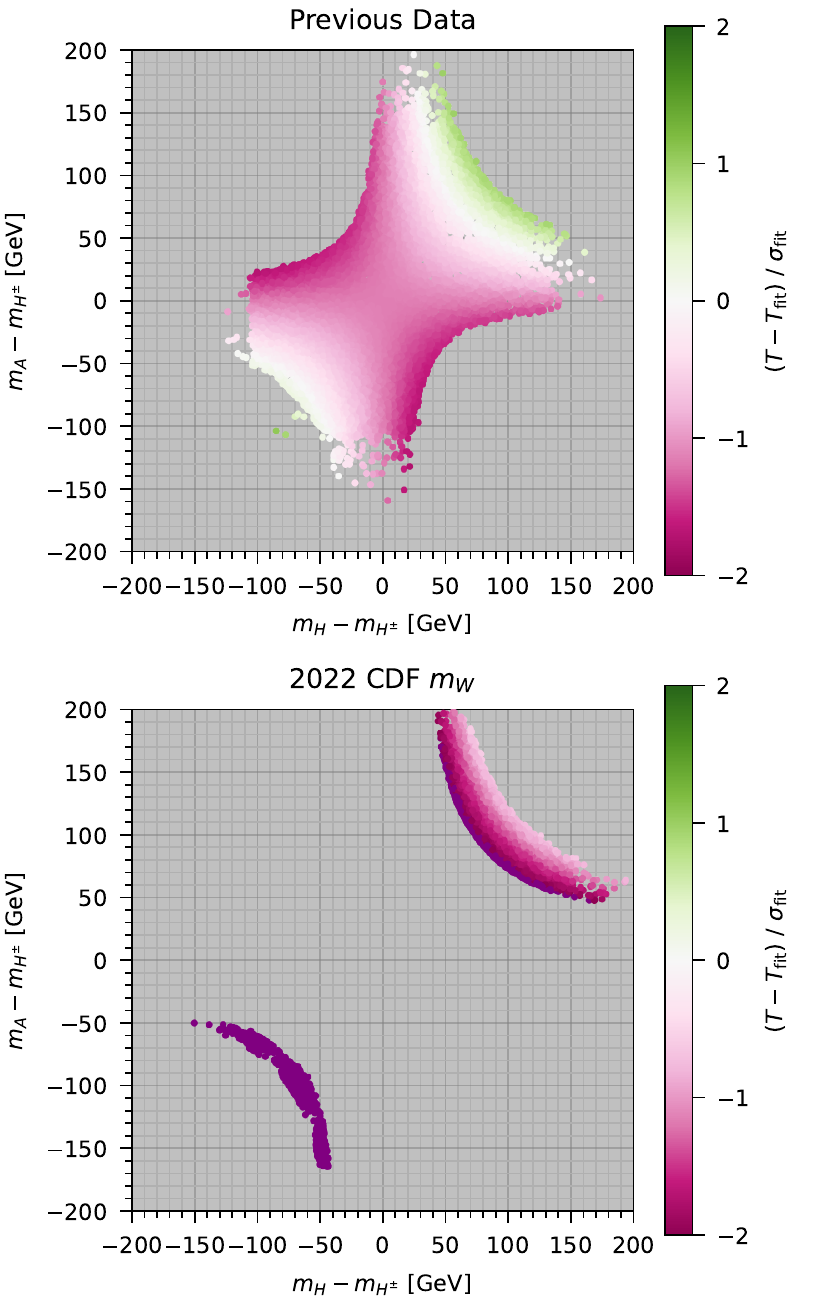}
    \caption{Likely mass separations between the charged Higgs boson and the two non-SM neutral Higgs bosons when $m_{H^\pm} = 100 \text{--} 170\,\text{GeV}$, in type-I 2HDM close to the alignment limit. $T_\text{fit}$ and $\sigma_\text{fit}$ in the upper(lower) plot are the central value and the uncertainty respectively of the fit of the $T$ parameter assuming $U = 0$\,\footnote{Type-I 2HDM predicts $\vert U \vert$ values no larger than 0.01 at all shown points.}, in Ref.~\cite{Haller2018}(Ref.~\cite{Lu2022}). All $m_H, m_A < m_{H^\pm}$ points in the lower plot have deviations exceeding $-2\sigma_\text{fit}$, which is possible since the electroweak precision constraints are a multivariate normal distribution.}
    \label{fig:mass-hierarchy}
\end{figure}
Before introducing the constraints on the parameter space of type-I 2HDM imposed by the $t\bar{t} \rightarrow W^+bb\bar{b}W^-\bar{b}$ decay, we note there are also various general theoretical and experimental results that especially implicate constraints on the mass space of the model. We use the \program{ScannerS-2} program \cite{ScannerS-2,ScannerS-1} to perform a scan on the parameter space defined as Table \ref{tab:param_scan} in type-I 2HDM. The scan implements constraints including:
\begin{itemize}
    \item Tree-level perturbative unitarity, boundedness from below, and absolute stability of tree-level electroweak vacuum.
    \item 95\% confidence level (CL) electroweak precision constraints, parametrized by the oblique parameters $S$, $T$, and $U$ \cite{Peskin1992}.
    \item 95\% CL flavor constraints from $b$-physics.
    \item 95\% CL constraints from searches for additional scalars and measurements of the 125\,GeV Higgs boson, realized by interfacing with \program{HiggsBounds 5.9.0} \cite{HiggsBounds_ref1, HiggsBounds_ref2, HiggsBounds_ref3, HiggsBounds_ref4, HiggsBounds_ref5} and \program{HiggsSignals 2.6.1} \cite{HiggsSignals_ref1, HiggsSignals_ref2}.
\end{itemize}
The electroweak precision constraints and the flavor constraints require a electroweak global fit, which will change significantly if we adopt the new $W$ boson mass reported by the CDF collaboration in 2022 \cite{CDF2022}. We use the results in Ref.~\cite{Haller2018} and Ref.~\cite{Lu2022} for the electroweak global fit of the previous data and the 2022 CDF data respectively. As of March 2023, the ATLAS collaboration reported a new $m_W$ value which is improved from while compatible with their previous data \cite{ATLAS2023}; we choose not to reconsider the fit for it. Further details of the implementation of the constraints and the scan can be found in Ref.~\cite{ScannerS-2}. We plot the constrained mass hierarchy of $H^+$, $A$, and $H$ for a light charged Higgs boson in Fig.~\ref{fig:mass-hierarchy}, which is in agreement with results from other scans \cite{Bahl2021, Lu2022, Ahn2022, Bahl2022}. The constraint from the $T$ parameter shows preference between two scenarios: (a) both $A$ and $H$ are heavier or lighter than the charged Higgs boson $H^+$, and (b) one of $A$ and $H$ is heavier than $H^+$ while the other one is lighter than $H^+$. Scenario (b), being in disfavor with the 2022 CDF data while likely allowed by previous data, describes the mass space studied in Section~\ref{sec:generic}.
\begin{table}[t]
    \caption{Ranges of the parameter scan. Masses are in GeV. $c(HVV)$ is the gauge coupling factor of the non-SM scalar $H$.}
    \label{tab:param_scan}
    \begin{ruledtabular}
    \begin{tabular}{ccccccc}
         & $m_{H^\pm}$ & $m_A$ & $m_H$ & $\tan\beta$ & $c(HVV)$ & $m_{12}^2$ \\
         \colrule
         min & 100 & 1 & 1 & 1 & $-0.1$ & 0 \\
         max & 170 & 300 & 300 & 25 & \phantom{+}0.1 & 250000 \\
    \end{tabular}
    \end{ruledtabular}
\end{table}

\begin{figure*}
    \begin{minipage}[t]{0.49\linewidth}
        \includegraphics{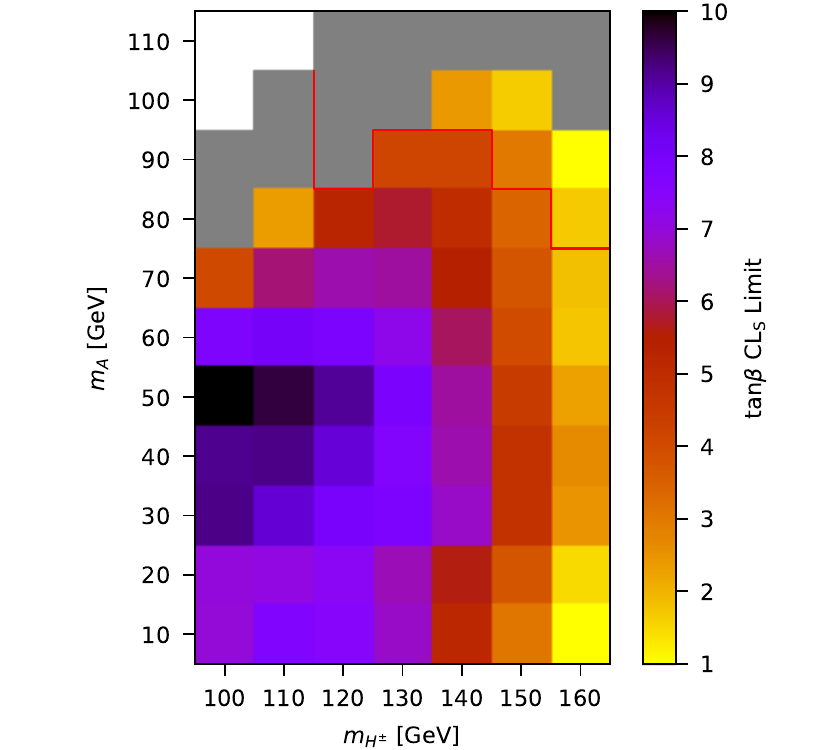}
        \caption{95\% $\text{CL}_\text{s}$ lower limits on $\tan\beta$ of type-I 2HDM on the $(m_{H^\pm}, m_A)$ plane, translated from Fig.~\ref{fig:BR_dist}. Each tile corresponds to a sample point which is located at the tile's center. $\tan\beta$ is not limited at grey-colored sample points, and limits at sample points above the red line are unreliable.}
        \label{fig:tanbeta_dist}
    \end{minipage}
    \hfill
    \begin{minipage}[t]{0.49\linewidth}
        \includegraphics{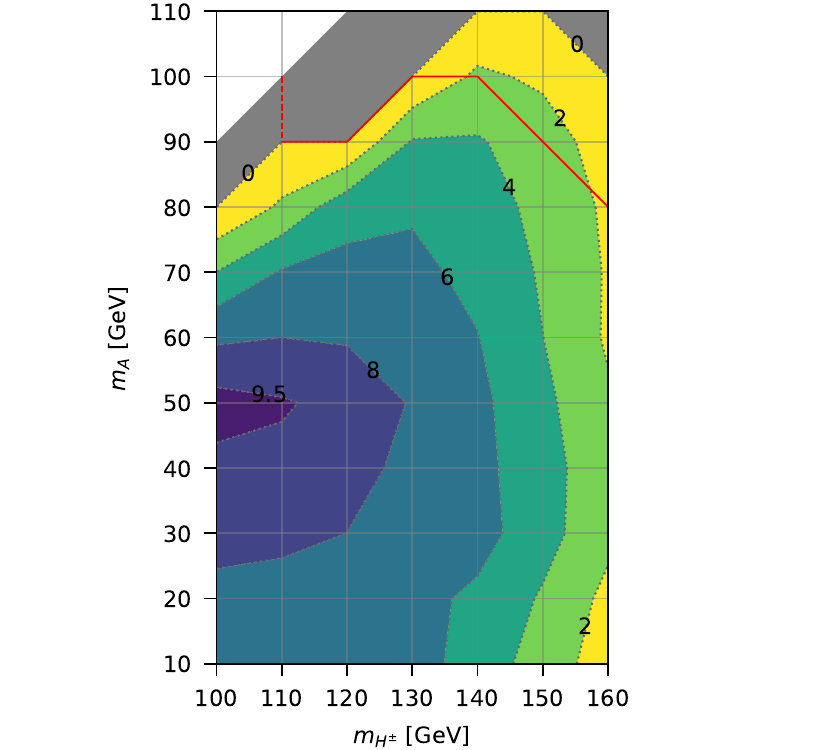}
        \caption{Intuitive contours of the 95\% $\text{CL}_\text{s}$ lower limits on $\tan\beta$ of type-I 2HDM on the $(m_{H^\pm}, m_A)$ plane. Limits at points above the red line are unreliable, except for those close to the dashed red line.}
        \label{fig:tanbeta_contours}
    \end{minipage}
\end{figure*}
\begin{figure*}
    \includegraphics{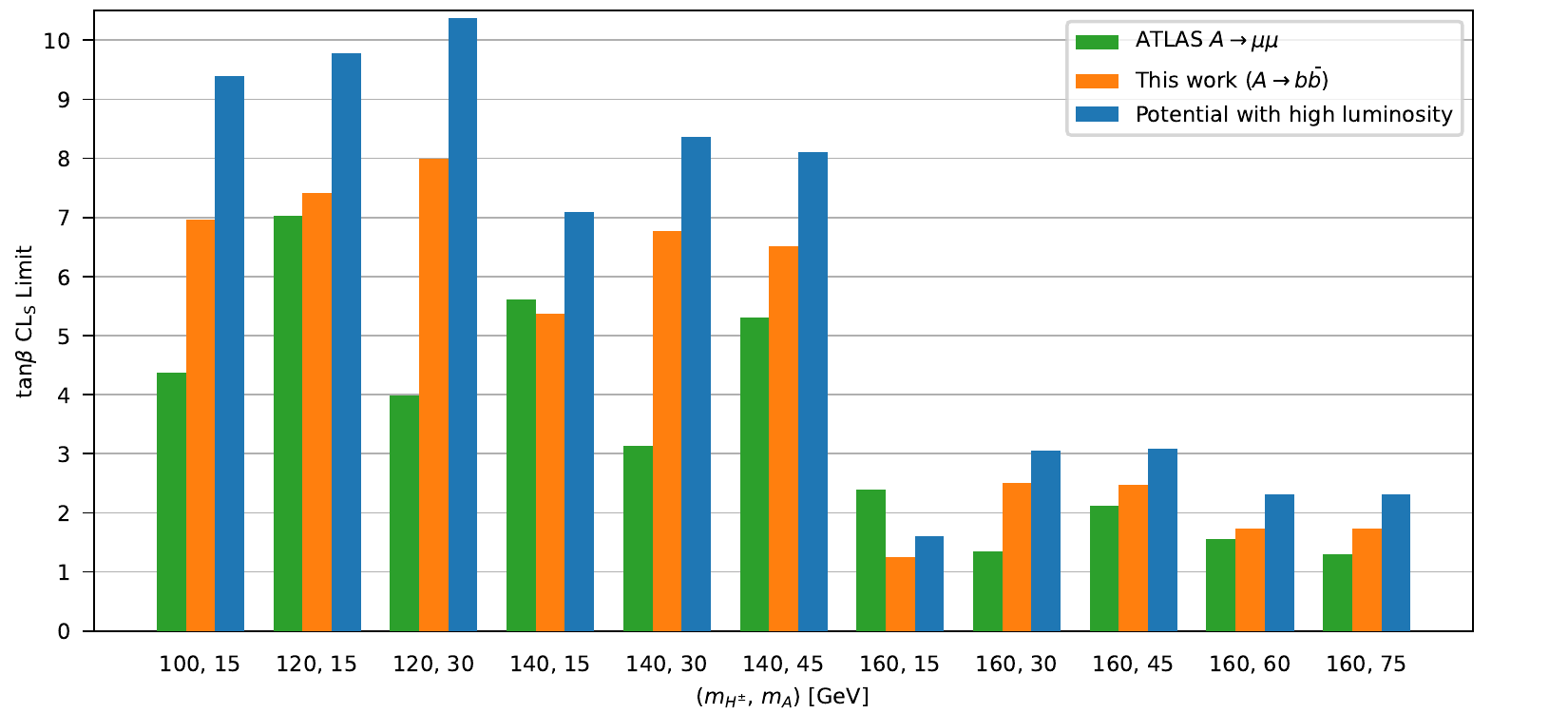}
    \caption{Comparisons between different results of 95\% $\text{CL}_\text{s}$ lower limits on $\tan\beta$ of type-I 2HDM. Comparisons are made at all simulated mass points listed in the ATLAS on-shell $H^\pm \rightarrow W^\pm A \rightarrow W^\pm \mu \mu$ study \cite{ATLAS2021_A_mumu}, and the limits in this work are linearly interpolated to get values at $m_A = 15,45,75\,\text{GeV}$. The "potential with high luminosity" result corresponds to a set of SM-like pseudodata lowering the current statistical uncertainties by 80\%, which is achievable in high-luminosity runs of the LHC.}
    \label{fig:comparison_tanbeta}
\end{figure*}
For type-I 2HDM with mass hierarchy $m_H > m_{H^\pm} > m_A$ and near the alignment limit, as is discussed following Eq.~(\ref{eqn:B_sig}) and Eq.~(\ref{eqn:param_space}) in the parameter space we consider: The signal strength $B^\text{sig}_{H^+}$ can be calculated as the product of three branching ratios, $B(t \rightarrow H^+ b)$, $B(H^+ \rightarrow W^{+(\ast)} A)$, and $B(A \rightarrow b \bar{b})$. The last branching ratio depends only on the mass of the pseudoscalar $A$, as all significant decay modes of $A$ originate from the couplings of $A$ to fermion pairs, which are all proportional to the mass of the fermion and $\cot\beta$ only. The product of the first two branching ratios can be roughly expressed from the related couplings as
\begin{eqnarray}
    &&B(t \rightarrow H^+ b) \times B(H^+ \rightarrow W^{+(\ast)} A)
    \nonumber\\ &\approx&
    [C_1(m_{H^\pm}, m_A) + C_2(m_{H^\pm}) \tan^2\beta]^{-1},
    \label{eqn:tanbeta_trend}
\end{eqnarray}
where $C_1$ and $C_2$ are positive constants, and $C_2$ increases as $m_{H^\pm}$ increases. Typical values of these two constants multiplied by $B(A \rightarrow b \bar{b})^{-1}$ are shown in Table~\ref{tab:trend_constants}.  Eq.~(\ref{eqn:tanbeta_trend}) approximates the trend about $\tan\beta$ quite well for $m_{H^\pm} - m_A \gtrapprox 40\,\text{GeV}$ at $\tan\beta \gtrapprox 1$. Therefore, we can deduce lower limits on $\tan\beta$ from the 95\% $\text{CL}_\text{s}$ upper limits on $B^\text{sig}_{H^+}$ for fixed masses of the Higgs bosons. We use the \program{ScannerS-2} \cite{ScannerS-2, ScannerS-1} program which is interfaced with \program{HDECAY 6.60} \cite{HDECAY_ref1, HDECAY_ref2, HDECAY_ref3} to calculate the branching ratios.
\begin{table}[t]
    \caption{Values of the two mass-related constants in $B^\text{sig}_{H^+} \approx (C_1' + C_2' \tan^2\beta)^{-1}$ at typical Higgs mass points ($m_H > m_{H^\pm}$).}
    \label{tab:trend_constants}
    \begin{ruledtabular}
    \begin{tabular}{rccccccc}
        $m_{H^\pm}$\,[GeV] & 100 & 100 & 140 & 140 & 160 & 160 & 160 \\
        $m_A$\,[GeV] & 20 & 50 & 20 & 80 & 50 & 80 & 120 \\
        \colrule
        $C_1'$\phantom{GeV} & 1.45 & 4.63 & 1.16 & 13.0 & 1.86 & 29.0 & 1917 \\
        $C_2'$\phantom{GeV} & 2.21 & 2.37 & 8.18 & 9.55 & 53.5 & 58.1 & 66.8 \\
    \end{tabular}
    \end{ruledtabular}
\end{table}

We plot the lower limits on $\tan\beta$ at the sample Higgs mass points in Fig.~\ref{fig:tanbeta_dist}, and the corresponding contours in Fig.~\ref{fig:tanbeta_contours}. The correlation of the $\tan\beta$ (or $\cot\beta$) limits with $m_A$ is attenuated at larger $m_{H^\pm}$ values. The most strict limit found among the sample points is 10.0, at $(m_{H^\pm} = 100\,\text{GeV}, m_A = 50\,\text{GeV})$, and the limit generally weakens as the masses of the two Higgs bosons move away in any direction within the mass space considered. We now note, however, we have ignored an alternative decay mode $H^+ \rightarrow t^{(\ast)} \bar{b} \rightarrow W^+ b \bar{b}$ which can make the result in Section~\ref{sec:generic} unreliable to be used here. We use two criteria to judge the significance of this alternative mode: (a) At the current limit value of $\tan\beta$, check if $B(H^+ \rightarrow t^{(\ast)} \bar{b}) > 0.1 \times B(H^+ \rightarrow W^+ A)$ (for reference, $B(A \rightarrow b \bar{b})$ is generally 90--70\% for $m_A = 10 \text{--} 110\,\text{GeV}$). (b) For points where $\tan\beta$ is not limited, check roughly if the maximum of the actual signal strength (with two decay modes summed up) can be larger than the previously calculated limit on $B^\text{sig}_{H^+}$. Points meeting the criteria are noted in Fig.~\ref{fig:tanbeta_dist} and linearly generalized in Fig.~\ref{fig:tanbeta_contours}. We also note that as too small $\tan\beta$ can violate the perturbative unitarity in the process studied in this work, our calculations do not include the region $\tan\beta < 1$, which is usually studied in other processes \cite{Sanyal2019}. The result expands our previous work \cite{prev_work} to the $(m_{H^\pm}, m_A)$ plane, showing a significantly more constrained area in $m_{H^\pm} \leqslant 130\,\text{GeV}$.

The most recent study of the $\tan\beta$ constraints for light $H^+$ and $A$ at various $(m_{H^\pm}, m_A)$, to the best of our knowledge, is included in a search for $H^\pm \rightarrow W^\pm A \rightarrow W^\pm \mu \mu$ where the $W$ boson is on shell, with the ATLAS detector \cite{ATLAS2021_A_mumu}. We compare our result (where $A \rightarrow b \bar{b}$) with this ATLAS result (where $A \rightarrow \mu \mu$) in Fig.~\ref{fig:comparison_tanbeta}. We find our constraints stronger than the constraints imposed by the $A \rightarrow \mu \mu$ mode at most Higgs mass points mutually included. The potential $\tan\beta$ limits from high-luminosity pseudodata which prefer SM within the current systematic uncertainties, translated from the potential signal limits in Fig.~\ref{fig:Projection_BR_contours}, are also compared in Fig.~\ref{fig:comparison_tanbeta}. At almost all sample points considered in this work, the potential limits on $\tan\beta$ are 21--38\% higher than the observed limits shown in Fig.~\ref{fig:tanbeta_dist}.

We can also consider the scenario where the masses of the pseudoscalar $A$ and the scalar $H$ are swapped. While the current $B^\text{sig}_{H^+}$ limits should still be applicable by simply replacing $A$ with $H$ as is discussed in Section~\ref{sec:generic}, the $\tan\beta$ limits can no longer be translated in the same way. The $H$-version $B^\text{sig}_{H^+}$ can depend differently on $\tan\beta$ and significantly on $m_{12}^2$: Differed from $A$, the non-SM scalar $H$ can decay into two photons through a charged-Higgs loop, and the related di-charged-Higgs coupling $c(H H^+ H^-)$ linearly depends on the soft-breaking $Z_2$ parameter $m_{12}^2$ and does not decrease to zero like the di-fermion couplings do as $\tan\beta \rightarrow \infty$ \cite{Branco2012, Arco2022}, making this di-photon channel dominant when $\tan\beta$ or $m_{12}^2$ is large. The constraints on the parameter space of type-I 2HDM in this scenario would be a completely different $\tan\beta \text{-} m_{12}^2$ distribution. We do not show this scenario here, as the constraints would be rather relaxed.

Another scenario involved is where both $A$ and $H$ are lighter than the charged Higgs boson. The $H$-part contribution to $B^\text{sig}_{H^+}$ is negligible in part of the parameter space, and in such cases, the previous $\tan\beta$ limits which are originally applied to $m_H > m_{H^\pm}$ can be reused, regardless of the presence of $H$. The branching ratios of the charged Higgs boson to $W^\pm A$ and $W^\pm H$ are almost equal near the alignment limit, while in the subsequent decay, $H \rightarrow b \bar{b}$ can be suppressed by three competing channels: $H \rightarrow Z A$, $H \rightarrow A A$, and the di-photon channel described in the previous scenario. The first two channels emerge only when $m_H > m_A$, whereas the di-photon channel can emerge at any $m_H$ as long as the soft-breaking $Z_2$ parameter $m_{12}^2$ is large enough for the limit value of $\tan\beta$. We note down an approximate criterion as an example for when the suppression by the di-photon channel can happen: For $m_{H^\pm} = 100 \text{--} 170\,\text{GeV}$, $m_H < m_A, m_{H_\text{SM}}, m_{H^\pm}$, and $\tan\beta \sim 10^0~\text{or}~10^1$, there is
\begin{eqnarray}
    \frac{m_H^{1.1}}{m_{H^\pm}^{2.1} \times 6\,\text{TeV}}
    \frac{m_{12}^2 \tan\beta}{\vert \tan(2\beta) \vert \sin(2\beta)}
    &\gtrapprox& 10
    \nonumber\\
    \Longleftrightarrow B(H \rightarrow b \bar{b}) &\lessapprox& 10\%.
\end{eqnarray}
We show this criterion on a moderately constrained point $(m_{H^\pm} = 140\,\text{GeV}, m_A = 60\,\text{GeV})$: The limit on $\tan\beta$ is 6.1 for $m_H > m_{H^\pm}$, so the formula gives $m_{12} > 0.61\,\text{TeV}$ for $m_H = 60\,\text{GeV}$. This range of $m_{12}$ actually corresponds to $B(H \rightarrow b \bar{b}) < 9.2\%$ if $\tan\beta$ is 6.1, while $B(A \rightarrow b \bar{b})$ is always 80\% at this mass point. Therefore, the limit on $\tan\beta$ will be almost identical to the previous value of 6.1 in this range of $m_{12}$. We also note that as the decay of $H$ into $A$ conversely suggests, the channel $A \rightarrow Z H$ emerges when $m_H < m_A$, and in such cases, the limit on $\tan\beta$ weakens and depends on $m_H$. A complete discussion of the $m_A, m_H < m_{H^\pm}$ scenario would involve too many free parameters, lying beyond our current method of reinterpreting the data.

\section{Summary}

We have studied the prospect of a light charged Higgs boson, which is produced from top quark pairs at the LHC, and decays into a $W$ boson and a pair of bottom quarks via an intermediate neutral Higgs boson. We set upper limits on the signal strength of this charged-Higgs channel with the ATLAS measurement at LHC 13\,TeV on the $WWbb\bar{b}\bar{b}$ final states, in which the distributions of the invariant mass of two closest $b$-jets show the greatest signal sensitivity. The 95\% $\text{CL}_\text{s}$ upper limit on the branching ratio $B(t \rightarrow H^+ b, H^+ \rightarrow W^+ H_i, H_i \rightarrow b \bar{b})$, where $H_i$ represents the neutral Higgs boson that participates in the decay, varies from 0.26\% to greater than 1.5\%, on the mass plane of a 100--160\,GeV charged Higgs boson and a 10--110\,GeV neutral Higgs boson. Other non-SM contributions to the same final states are not included, yet they are insignificant for most Higgs masses sampled if considered in 2HDM. The limits are expected to be lowered by 32--45\% for most Higgs masses sampled, with future high-luminosity data if SM is preferred then.

The signal limits are translated into constraints on the parameter space of type-I 2HDM, where we have especially discussed the current general constraints on the possible hierarchies of the Higgs masses. We discuss the parameter constraints with specific mass hierarchies, as we argue the decay properties of the $CP$-odd Higgs boson $A$ and the $CP$-even Higgs boson $H$ are different. The 95\% $\text{CL}_\text{s}$ lower limit on $\tan\beta$ when $m_H > m_{H^\pm} > m_A$ varies from 1 to 10; future high-luminosity data can potentially raise the limits by 21--38\% for most Higgs masses sampled. The result demonstrates the power of the $W^\pm b \bar{b}$ final states of the charged Higgs boson in constraining the parameter space of type-I 2HDM or models with similar couplings. We encourage dedicated experimental searches for further improvements.

%\phantom{}
\begin{acknowledgments}

The work of J. G. is supported by the National Natural Science Foundation of China under Grants No. 12275173 and No. 11835005.
%We realized a special implementation of interpolation of smooth plane curves
%(see the repository \mbox{February-L}/Uniform-Acceleration-Interpolation on
%\textit{GitHub}); it is eventually unused but may be helpful for data
%presentation in other studies.

\end{acknowledgments}

\bibliography{Hc-II_ref}

\end{document}